# Connection between charge-density-wave order and charge transport in the cuprate superconductors

W. Tabis[1,2,+], Y. Li[3,4], M. Le Tacon[5], L Braicovich[6], A. Kreyssig[7], M. Minola[5], G. Dellea[6], E. Weschke[8], M. J. Veit[1], M. Ramazanoglu[7], A. I. Goldman[7], T. Schmitt[9], G. Ghiringhelli[6], N. Barišić[1,10,11], M. K. Chan[1], C. J. Dorow[1], G. Yu[1], X. Zhao[1,12], B. Keimer[5] and M. Greven[1]

[1] School of Physics and Astronomy, University of Minnesota, Minneapolis, Minnesota 55455, USA

[2] AGH University of Science and Technology, Faculty of Physics and Applied Computer Science, 30-059 Krakow, Poland

[3] International Center for Quantum Materials, School of Physics, Peking University, Beijing 100871, China

[4] Collaborative Innovation Center of Quantum Matter, Beijing 100871, China

[5] Max Planck Institute for Solid State Research, D-70569 Stuttgart, Germany

[6] CNR-SPIN, CNISM and Dipartimento di Fisica, Politecnico di Milano, I-20133 Milano, Italy

[7] Ames Laboratory and Department of Physics and Astronomy, Iowa State University, Ames, Iowa 50011, USA

[8] Helmholtz-Zentrum Berlin für Materialien und Energie, Albert-Einstein-Straße 15, D-12489 Berlin, Germany

[9] Research Department Synchrotron Radiation and Nanotechnology, Paul Scherrer Institut, CH-5232 Villigen PSI, Switzerland

[10] Service de Physique de l'Etat Condensé, CEA-DSM-IRAMIS, F 91198 Gif-sur-Yvette, France

[11] Institute of Solid State Physics, Vienna University of Technology, 1040 Vienna, Austria

[12] State Key Lab of Inorganic Synthesis and Preparative Chemistry, College of Chemistry, Jilin University, Changchun 130012, China

[+] Current address: Laboratoire National des Champs Magnétiques Intenses, Toulouse 31400, France

**Charge-density-wave (CDW) correlations within the quintessential CuO$_2$ planes have been argued to either cause [1] or compete with [2] the superconductivity in the cuprates, and they might furthermore drive the Fermi-surface reconstruction in high magnetic fields implied by quantum oscillation (QO) experiments for YBa$_2$Cu$_3$O$_{6+\delta}$ (YBCO) [3] and HgBa$_2$CuO$_{4+\delta}$ (Hg1201) [4]. Consequently, the observation of bulk CDW order in YBCO was a significant development [5,6,7]. Hg1201 features particularly high structural symmetry and recently has been demonstrated to exhibit Fermi-liquid charge transport in the relevant temperature-doping range of the phase diagram, whereas for YBCO and other cuprates this underlying property of the CuO$_2$ planes is partially or fully masked [8-10]. It therefore is imperative to establish if the pristine transport behavior of Hg1201 is compatible with CDW order. Here we investigate Hg1201 ($T_c$ = 72 K) via bulk Cu *L*-edge resonant X-ray scattering. We indeed observe CDW correlations in the absence of a magnetic field, although the correlations and competition with superconductivity are weaker than in YBCO. Interestingly, at the measured hole-doping level, both the short-range CDW and Fermi-liquid transport appear below the same temperature of about 200 K. Our result points to a unifying picture in which the CDW formation is preceded at the higher pseudogap temperature by q = 0 magnetic order [11,12] and the build-up of significant dynamic antiferromagnetic correlations [13]. Furthermore, the smaller CDW modulation wave vector observed for Hg1201 is consistent with the larger electron pocket implied by both QO [4] and Hall-effect [14] measurements, which suggests that CDW correlations are indeed responsible for the low-temperature QO phenomenon.**

Figures 1a,b show the combined temperature-doping phase diagram of Hg1201 and YBCO, which features an insulating state with antiferromagnetic (AF) order at low hole-dopant concentrations, unusual translational-symmetry-preserving (**q** = 0) magnetism [11,12] below the pseudogap temperature $T^*$, a Fermi-liquid regime below $T^{**}$ [8-10], and an approximately parabolic "superconducting dome" $T_c(p)$ [15,16]. The CDW in YBCO has been reported below optimal doping, for hole concentrations that correspond to the 'plateau' in $T_c(p)$, [6,7,17,18] (Fig. 1b,c).

We report bulk resonant X-ray scattering measurements of Hg1201 ($p \approx 0.09$, $T_c = 72$ K), close to the center of its $T_c(p)$ plateau. Hg1201 features a simple tetragonal crystal structure, the highest optimal superconducting transition temperature ($T_{c,max} \approx 97$ K) of all cuprates with one CuO$_2$ layer per primitive cell [19] (see also supplementary information (SI)). For $p \approx 0.09$, Hg1201 exhibits Fermi-liquid charge transport in zero [8,9] and moderate [10] magnetic field at intermediate temperatures, as well as Shubnikov-de-Haas oscillations in high magnetic fields at temperatures well below the zero-field $T_c$ [4]. For the present experiments, single crystals were grown and prepared according to previously described procedures [19,20].

Resonant X-ray scattering techniques combine the sensitivity of traditional X-ray diffraction to ordered structures with the advantages of site-selective X-ray absorption spectroscopy. By tuning the incident photon energy to match the Cu $L_3$-edge, the CuO$_2$ layers can be studied directly through excitations of core $2p_{3/2}$ electrons into the Cu $3d$ valence band, with a high sensitivity to spatial charge modulations. Resonant X-ray diffraction (RXD) allows the observation of order of different electronic states of the same atomic species in a solid. For Cu $L_3$-edge RXD, the spatial charge modulation of the Cu $3d$ states is reflected in a modulation of the (complex) scattering factor, giving rise to the superstructure peaks at momenta corresponding to the real space periodicity. Moreover, resonant inelastic X-ray scattering (RIXS) offers the additional possibility to resolve the energy of the scattered photons, thus allowing the separation of the CDW contribution from phenomena at higher energy transfers.

We quote the scattering wave vector as $\mathbf{Q} = H\mathbf{a}^* + K\mathbf{b}^* + L\mathbf{c}^*$, where $\mathbf{a}^*$, $\mathbf{b}^*$ and $\mathbf{c}^*$ are reciprocal lattice vectors. Similar to prior work [6,17], the relatively small energy of the soft X-rays constrained us to work in the first Brillouin zone. The measurements were furthermore constrained to scans of the type [$H$,0,$L$], in which $L$ was dependent on the planar momentum transfer $H$ (see also SI).

Figure 2 summarizes our observation of CDW correlations in Hg1201 via both RXD (Fig. 2a,b,c) and RIXS (Fig. 2e,f). RXD data were collected at the Cu $L_3$-resonance (931.7 eV) with the incident beam polarized perpendicular to the scattering plane (σ-polarization). The momentum scan at 70 K (Fig. 2a) displays a pronounced peak that is absent away from resonance (Fig. 2b), which proves the involvement of the electronic degrees of freedom in the scattering process, without contribution from atomic displacements. The scan at 250 K demonstrates that CDW order is absent at this temperature (Fig. 2b), and only a temperature-independent, non-monotonic background is observed. The peak position, $H_{CDW} = 0.276(5)$ r.lu., and FWHM width, $2\kappa = 0.058(9)$ r.l.u., at 70 K are best extracted from the intensity difference between the two temperatures (Fig. 2c).

This finding is consistent with our RIXS experiment performed with σ-polarization and an energy resolution of 130 meV. This technique allowed the separation of the quasi-elastic peak from magnetic, $3d$ inter-orbital and O $2p$/Cu $3d$ charge-transfer excitations at higher energy (Fig. 2d). As demonstrated in ref. [6], the quasi-elastic peak contains information about the spatial charge modulations. Figure 2d shows the low-energy inelastic spectra at three momenta for both 75 K and 220 K. There is an enhancement due to CDW formation at 75 K. The full $H$ dependence is shown in Fig. 2e. From a Gaussian fit, the peak position, $H_{CDW} = 0.280(5)$ r.l.u., and FWHM width, $2\kappa = 0.071(3)$ r.l.u., are in very good agreement with the RXD experiment.

Figure 1e shows the temperature dependence of the CDW peak intensity, obtained with the two X-ray techniques, indicating that CDW order is observed below $T_{CDW} = 200(15)$ K. As further demonstrated in the figure, the CDW phenomenon occurs about 100 K below the onset of **q** = 0 order [11,12], which in turn coincides with $T^*$ obtained from resistivity [8,10]. For YBCO, $T_{CDW}$ falls in a rather narrow temperature range (135-160 K; see Fig. 1) [6,7,17,18]. When combined with our result for Hg1201, it becomes apparent that $T_{CDW}$ is consistent with $T^{**}$, the temperature at which the Seebeck coefficient reaches its maximum value [15,21], and below which conventional Fermi-liquid planar charge transport is observed [8-10] (see Fig. 1a,c). Recent magneto-resistance measurements revealed for tetragonal Hg1201 ($p \approx 0.09$ and 0.11) that Kohler's rule is obeyed below $T^{**}$, with a Fermi-liquid quasiparticle scattering rate, in agreement with Boltzmann transport theory [10]. For YBCO, due to lower (orthorhombic) symmetry and the presence of Cu-O chains (along the $b$–axis) that reside between the CuO$_2$ planes, Fermi-liquid transport is revealed only in de-twinned samples, for current flow perpendicular to the chains [8,10]. At least for Hg1201, the arc-like "Fermi-surface" (FS) should therefore be temperature-independent in an extended temperature range below $T^{**}$. Since the underlying Fermi-liquid regime appears to extend to very low hole concentrations [8], whereas the CDW order appears to be tied to the $T_c(p)$ plateau (Fig. 1d shows that $\xi_{CDW}$ for YBCO decreases away from $p = 0.12$), the stable FS implied by the transport measurements for Hg1201 likely is a necessary, but not sufficient condition for the occurrence of CDW correlations.

The CDW order in our Hg1201 sample is rather weak and not significantly impacted by the onset of superconductivity: the planar correlation length of $\xi_{CDW} = a/(2\pi\kappa) \approx 5a$ (inset to Fig.

1e) is temperature-independent and only slightly larger than the lower bound of $1/H_{CDW} \approx 3.5a$ set by the modulation period. Furthermore, in contrast to YBCO [7], we are unable to identify CDW correlations via (non-resonant) hard X-ray diffraction (see SI). For YBCO ($p = 0.11$ and 0.13 [6]), $\xi_{CDW}$ takes on a similar value at high temperature, but is found to increase upon cooling toward $T_c$: Fig. 1d shows $\xi_{CDW}$ both at $T_c$ and the extrapolated value at $T_{CDW}$. Interestingly, $\xi_{CDW}(p)$ takes on a maximum at $p = 0.12$, where the deviation of the $T_c(p)$ from a simple parabolic shape is greatest for YBCO (Fig. 1b,c). This deviation appears to reflect the competition between CDW order and superconductivity [2,6,7]. If the magnitude of the deviation is taken as a measure of this competition, then the CDW correlations would indeed be expected to be weaker in Hg1201. This is supported by the fact that, in YBCO, the strongest suppression of the CDW amplitude in the superconducting state was observed for $p = 0.12$ [6]. In contrast, we find no decrease below $T_c$ for Hg1201.

Given the myriad unconventional phenomena exhibited by the cuprates, especially the peculiar high-temperature $\rho \sim T$ metallic behavior and the opening of a pseudogap along certain portions of the FS at temperatures below $T^*$, it came as a surprise that moderately-doped YBCO exhibits quantum oscillations [3], the hallmark of a Fermi-liquid metal with a coherent FS. CDW correlations are considered a likely candidate for the FS reconstruction implied by the QO experiments. NMR experiments indicate that a high magnetic field induces static long-range CDW order in YBCO [5]. The Hall, Nernst and Seebeck coefficients of YBCO and Hg1201 exhibit remarkably similar high-magnetic-field behavior, including a low-temperature sign change, which suggests the possible existence of electron pockets as a result of the FS reconstruction [14]. The observation of QO in Hg1201 proved that pockets are indeed universally present, and not the result of the structural complexity of YBCO [4].

Figure 3a (Fig. 3b) demonstrates that the CDW modulation wave vector (QO period $F$) is significantly smaller (larger) for Hg1201 than for YBCO [6,7,17,18,22] at comparable doping levels. If we treat the hole concentration as an intrinsic parameter, we find an approximately linear relationship between $F$ and $H_{CDW}$ for the combined data (Fig. 3c). Figure 3d shows a fit to a plausible quadratic dependence of the fractional pocket size on $H_{CDW}$, as would be expected in the simple picture of a reconstructed nodal electron pocket due to biaxial CDW order that involves segments of the underlying Fermi surface defined by $H_{CDW}$ [23]. This is shown schematically in the inset to Fig. 3d, where tight-binding parameters have been used to estimate

the underlying FS of Hg1201 [24] and YBCO (bonding FS) [7,23]. For Hg1201, the same tight-binding parameters were recently found to be in reasonably good agreement with photoemission measurements of the nodal FS closer to optimal doping [24]. Interestingly, the so-obtained pocket sizes agree with the values of the QO experiments, as shown in Fig. 3d. They furthermore agree with the relative values of the (negative) Hall coefficients obtained in the low-temperature, high-magnetic-field limit [14]. These observations suggest that the QO phenomenon is indeed a consequence of the universal CDW correlations. The Fermi pocket reconstructed from nodal Fermi arcs is consistent with checkerboard-type CDW order [25] and inconsistent with unidirectional charge-stripe order [26]. Our data do not allow us to test recent theoretical predictions that involve hotspots [27,28] or twice the anti-nodal distance [29], since the optimal CDW wave vector does not have to connect the hotspots, and because underlying FS far from the nodal direction (in the absence of the PG and CDW order) are unknown.

Recent combined RXD and scanning-tunneling microscopy (STM) work demonstrated the presence of CDW correlations in both $Bi_2Sr_{2-x}La_xCuO_{6+\delta}$ (Bi2201) [27] and $Bi_2Sr_2CaCu_2O_{8+\delta}$ (Bi2212) [30], with values of $H_{CDW}$ between 0.25 to 0.30 r.l.u., and consistent between the bulk (RXD) and surface (STM) techniques. Similar to the present result for Hg1201, $\xi_{CDW}$ was found to be rather small (~7$a$). Whereas YBCO exhibits a clear anti-correlation between CDW and SC order, for single-layer Bi2201 and double-layer Bi2212 the RXD signal is not significantly suppressed below $T_c$ [27]. All three cuprates exhibit lower structural symmetry than Hg1201, and therefore are less ideal for a quantitative study of the connection between charge transport and CDW order. Although QO below optimal doping were first discovered in YBCO [3], pristine planar Fermi-liquid transport ($\rho \sim T^2$ in zero magnetic field; confirmed by the validity of Kohler's rule for the magnetoresistance) is only found for current flow perpendicular to the Cu-O chains. Bi2201 and Bi2212 are more disordered [8,31], which has prevented the observation of Fermi-liquid transport, including QO. For Bi2201, a direct correlation between $H_{CDW}$ and the distance between the Fermi arc tips was suggested based on combined photoemission and X-ray work [27]. Whereas Ref. [30] suggests that PG formation precedes CDW order in Bi2212, consistent with our finding, it is argued in Ref. [27] that $T_{CDW} \approx T^*$ for Bi2201. However, given the statistical uncertainty in the RXD data for Bi2201, the appearance of CDW order may also be interpreted to coincide with $T^{**}$ [27].

The present work provides strong support for the idea that incommensurate CDW correlations drive the FS reconstruction implied by the QO experiments. Moreover, it confirms that much can be learned from a comparative study of Hg1201 with other cuprates, especially with YBCO. Neutron scattering experiments have demonstrated universal **q** = 0 magnetism [11] along with a significant build-up of gapped AF correlations [13] upon the formation of the pseudogap at $T^*$. The combined results point to the distinct possibility that the sequence of ordering tendencies (**q** = 0 order precedes CDW order, which in turn precedes SC order) and the phase diagram as a whole are driven by AF correlations [26,28,32].

**Author Information:** Correspondence and requests for materials should be addressed to W.T. (tabis@physics.umn.edu ) and M.G. (greven@physics.umn.edu)

**Acknowledgements:** We thank C. Proust, B. Vignolle, and D. Vignolles for useful discussions. We thank V.N. Strocov and V. Bisogni for technical and user support at SLS, E. Schierle at BESSY-II, and D. Robinson at APS. The work at the University of Minnesota was supported by the U.S. Department of Energy, Office of Basic Energy Sciences. N.B. acknowledges support though a Marie Curie Fellowship and the European Research Council. AK and AIG were supported by the U.S. Department of Energy, Office of Basic Energy Sciences. Ames Laboratory is operated for the U.S. Department of Energy by Iowa State University.

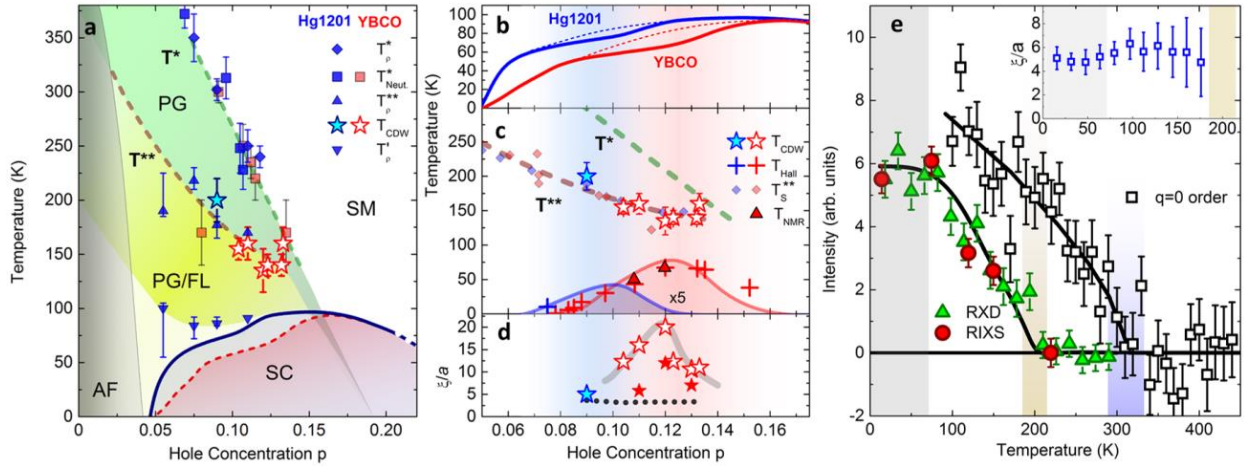

**Figure 1 | Phase diagram and CDW versus q=0 order. a,** Joint phase diagram of Hg1201 and YBCO. Solid blue (dashed red) line: superconducting dome of Hg1201 [15] and YBCO [16]). Blue (red) symbols correspond to characteristic temperatures of Hg1201 (YBCO): Pseudogap (PG) temperature $T^*$, determined from deviation from linear-$T$ resistivity in the "strange" metallic (SM) regime [8] and from neutron scattering experiments of the onset of $q = 0$ magnetism [11,12]. Yellow area: Fermi-liquid (FL) regime between $T^{**}$ and the measureable onset of SC fluctuations at $T_\rho'$ [8,10]. Grey shaded area: approximate extent of antiferromagnetic (AF) phase for YBCO. Stars indicate $T_{CDW}$. **b,** Solid lines: $T_c(p)$ [15,16]. Dashed lines: hypothetical 'ideal' (approximately parabolic) "superconducting domes." **c,** Solid lines: deviation of the SC dome from ideal shape (multiplied by 5). The hole concentration $p$ needs to be viewed as an effective parameter, with $p = 0.09$ for Hg1201 corresponding to $p = 0.11$-$0.12$ for YBCO. Crosses: temperatures at which the Hall coefficient changes sign in a high magnetic field [14]. This sign change is attributed to FS reconstruction. Red triangles: onset of (nearly) static charge order in YBCO, observed by NMR in a high magnetic field [5]. Dynamic CDW correlations appear below $T_{CDW}$, consistent with $T^{**}$ obtained from planar resistivity ($T_\rho^{**}$) [8] and Seebeck ($T_S^{**}$) (Hg1201 [14]; YBCO [21]) measurements. **d,** CDW correlation length (in units of the planar lattice constant $a$) for YBCO [6,7,17,18] and Hg1201. The open stars represent the correlation length at $T_c$ and the solid stars the correlation length just below $T_{CDW}$. The maximum correlation length for YBCO is observed at $p \approx 0.12$. Dotted line represents the CDW modulation period $1/H_{CDW}$. Green and brown dashed lines in **a** and **c** indicate $T^*(p)$ and $T^{**}(p)$, respectively, and are guides to the eye. **e,** Temperature dependence of the CDW intensity (integrated along $H$)

for Hg1201 obtained by RIXS (red circles) and RXD (green triangles). The CDW emerges below $T_{CDW}$ = 200(15) K, its intensity increases on cooling toward $T_c$ = 72 K, and then levels off (shaded grey indicates SC phase). The CDW correlations appear well below the opening of the pseudogap at $T^* \approx$ 320 K which is associated with $\mathbf{q}$ = 0 magnetic order (squares; order parameter data for a $T_c$ = 75 K sample [12]), a deviation from $T$-linear planar resistivity, and a considerable build-up in AF correlations [13]. Inset: Absence of temperature dependence of the CDW correlation length. Vertical blue and brown bands indicate $T^*$ and $T^{**}$, respectively.

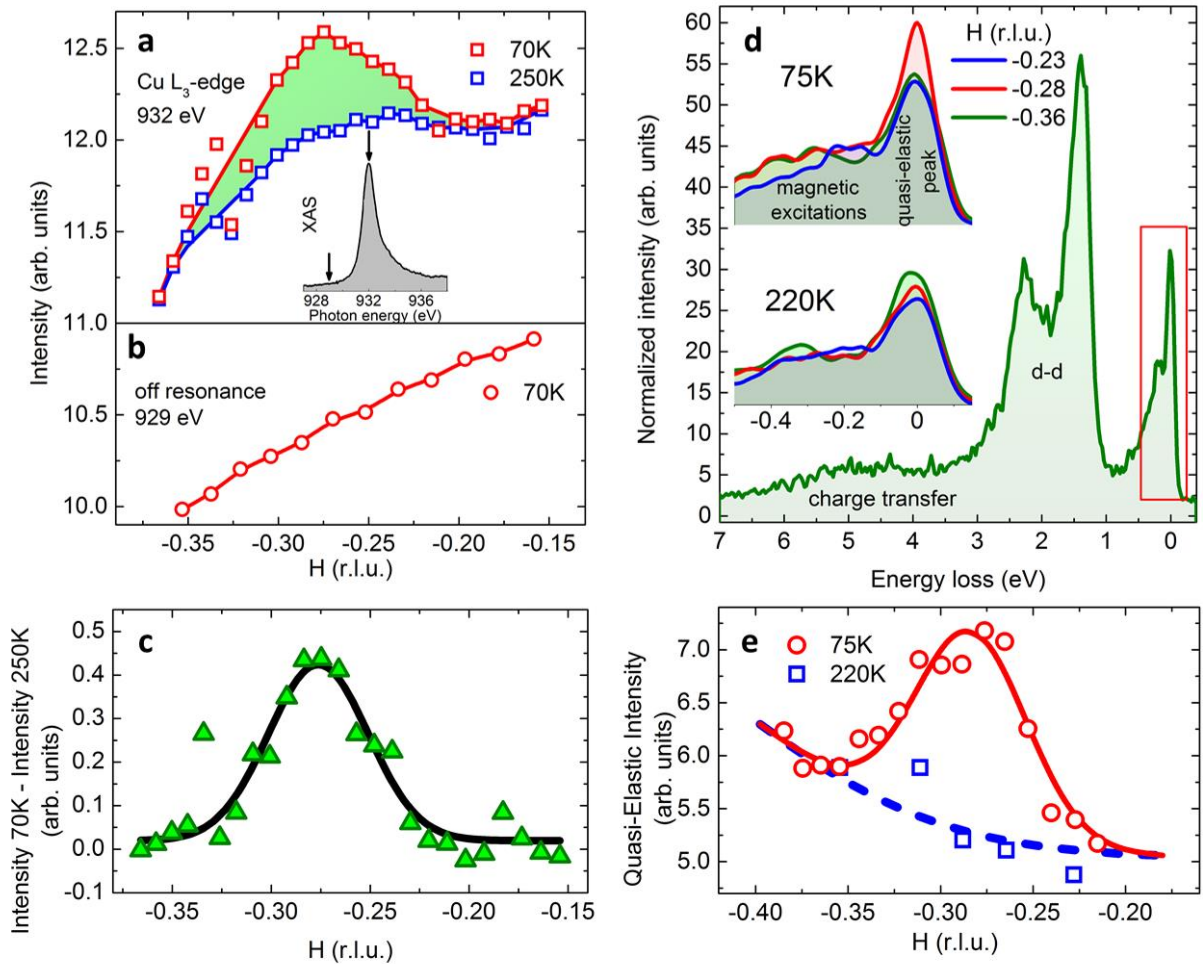

**Figure 2 | CDW in Hg1201 ($T_c$ = 71 K) observed with RXD and RIXS. a,** RXD spectra collected around the CDW propagation vector at Cu $L_3$-resonance at 70 K and 250 K. At resonance, the scattering factor becomes complex ($f_i' + if_i''$) and depends strongly on the valence state of the resonating Cu atoms. CDW correlations give rise to the additional intensity observed at 70 K (red squares). The non-monotonic background does not change at temperatures above 200 K. The inset shows the total fluorescence X-ray absorption spectrum at 70 K. Arrows indicate the energies at which the momentum scans in **a** and **b** were performed. **b,** The CDW peak at 70 K is not observed out of resonance. **c,** Intensity difference between scans at 70 K and 250 K and fit to a Gaussian function (solid line). **d,** RIXS spectra at the Cu $L_3$-edge (931.7 eV) obtained with the polarization vector perpendicular to the scattering plane (σ-polarization). Excitations typical of the cuprates are identified. The insets are zooms (red frame) for 75 K and 220 K at three values of **Q** around the CDW wave vector. The enhancement of the quasi-elastic peak due to the CDW is observed for $H_{CDW} \approx 0.28$ r.l.u. The spectra are normalized to the *d-d*

excitation intensity. In contrast to prior Cu *K*-edge RIXS measurements of charge-transfer excitations in Hg1201 [33] (where the 4*p* final state serves as a 'spectator' and the valence band is probed indirectly through the Coulomb potential of the core hole and of the excited electron), the present experiment at the Cu $L_3$-edge is directly sensitive to the valence band. **e,** Integrated intensity of the quasi-elastic peak as a function of *H* at 75 K and 220 K. Data are fit to a Gaussian peak with parabolic background. The peak widths in **c** and **e** correspond to correlation length values of $\xi_{CDW}/a = 5.5(8)$ and $4.6(2)$, respectively.

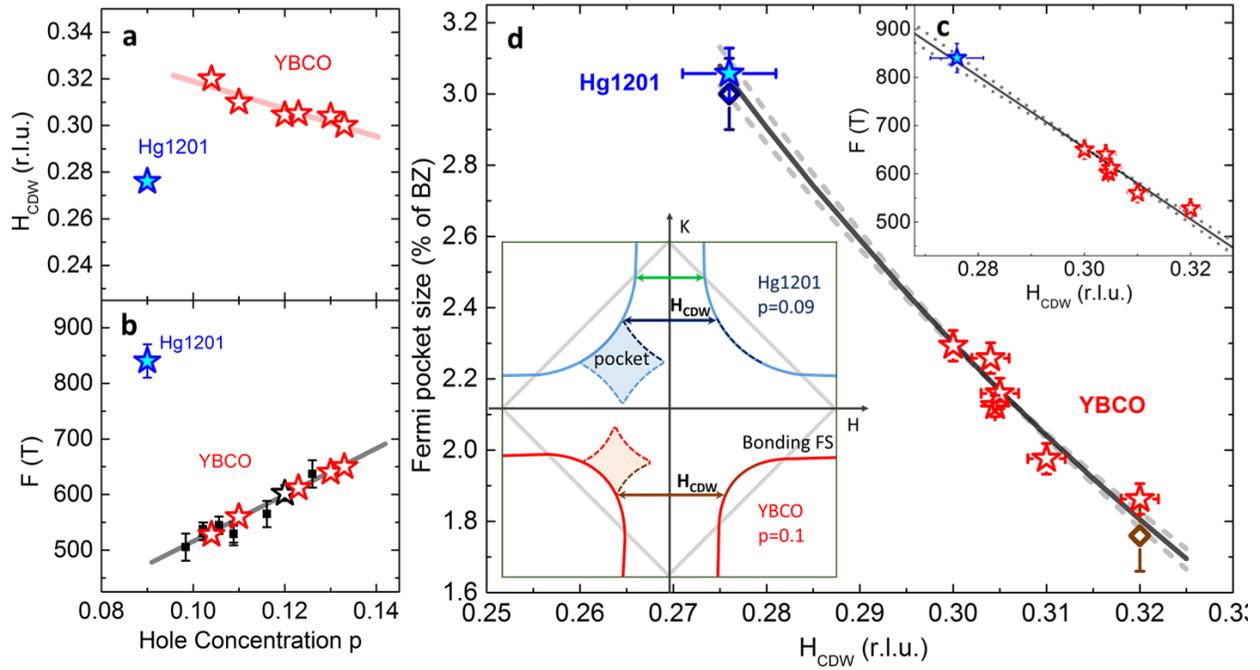

**Figure 3 | Relationship between CDW wave vector and reconstructed Fermi pockets. a**, $H_{CDW}$ for Hg1201 (blue star) and YBCO (red stars; estimated from Refs. [6,7,17,18]) as a function of hole concentration. **b**, Quantum oscillation frequency $F$ for Hg1201 (blue star) [4] and YBCO (black squares, black star) [3,22] as a function of hole concentration. For Hg1201, the X-ray and QO data were obtained for samples prepared under the same conditions. The black star for YBCO indicates a doping level for which both CDW and QO were measured. Red stars represent the QO frequencies estimated form the linear fit for the doping levels at which only CDW order (but not QO) was studied. **c**, QO frequency versus $H_{CDW}$ for Hg1201 (blue) and YBCO (red), with doping as intrinsic parameter. The solid line represents a linear fit. The dashed lines show the fit error range. **d**, Fermi pocket size as a fraction of the first Brillouin zone (BZ) versus $H_{CDW}$ for Hg1201 (blue) and YBCO (red), with doping as intrinsic parameter. The solid line represents a quadratic fit, which extrapolates to $H_{CDW} = 0.45(2)$ r.l.u. in the limit of vanishing pocket size (see also SI). The dashed lines show the fit error range. Inset: Schematic of reconstructed nodal electron pocket defined by $H_{CDW}$ and by the simple tight-binding Fermi surface for both Hg1201 ($p = 0.09$ [24]) and YBCO (bonding band; $p = 0.10$, obtained from taking the average for $p = 0.12$ [7] and $p = 0.08$ [23]), with BZ (green square) and with AF BZ boundary (grey line). The respective pocket areas are $3 \pm 0.1\%$ and $1.8 \pm 0.1\%$ and are consistent

with experiment, as indicated by dark blue and brown diamonds in **d**. The green double-arrow connects the 'hotspots' of the tight-binding Fermi surface of Hg1201 and is substantially smaller (~0.21 r.l.u.) than the measured value $H_{CDW} \approx 0.28$ r.l.u.

# Connection between charge-density-wave order and charge transport in the cuprate superconductors

W. Tabis, Y. Li, M. Le Tacon, L Braicovich, A. Kreyssig, M. Minola, G. Dellea, E. Weschke, M. J. Veit, M. Ramazanoglu, A.I. Goldman, T. Schmitt, G. Ghiringhelli, N. Barišić, M. K. Chan, C. J. Dorow, G. Yu, X. Zhao, B. Keimer and M. Greven

Supplementary information

**Crystal structures of Hg1201 and YBCO**

The structures of Hg1201 and YBCO exhibit considerable differences (Fig. S1). Hg1201 features tetragonal symmetry (space group P4/mmm) and one $CuO_2$ plane per primitive cell. In the SC doping range, YBCO features lower, orthorhombic symmetry (space group P/mmm) and two $CuO_2$ planes per primitive cell. In both compounds, the charge carrier concentration in the $CuO_2$ planes is controlled by the interstitial O concentration. In YBCO, these O atoms form Cu-O chains along the *b*-axis, with various inter-chain ordering patterns [16]. The interstitial O atoms in Hg1201 reside in the Hg-O planes and do not order in the doping range of interest.

**Sample preparation**

Hg1201 single crystals were grown by a two-step flux method [20], subsequently annealed at 400°C in partial vacuum of 100 mtorr [19], and then quenched to room temperature. The width of the superconducting transition, as determined from magnetic susceptibility, was typically 2 K with a midpoint of $T_c = 72$ K. According to the $T_c$ vs. $p$ relationship established for polycrystalline samples [15], this corresponds to a hole concentration of $p \approx 0.09$. The typical dimensions of the crystals in this study were 2 x 2 x 0.5 mm$^3$. The X-ray penetration length in Hg1201 is about 0.2 μm at the energy used in the soft X-ray experiments ($h\omega \approx 932$ eV). To ensure a flat and clean surface, each sample was polished multiple times with sandpaper with increasingly finer grade, ranging from 1 μ m to 0.05 μ m. The same sample preparation process was used in separate X-ray absorption spectroscopy measurements of Hg1201 at O *K* and Cu *L*[3]- edges, giving us high confidence that the procedure does not change the doped hole concentration.

**Experimental setup**

RXD measurements at the Cu $L_3$ – edge were performed at UE46 beam line of the BESSY-II synchrotron in Berlin, Germany [35]. In order to determine the exact resonance energy, X-ray absorption measurements were performed in a fluorescence configuration. The incident X-ray beam was then tuned to the maximum of the fluorescence signal at $h\omega$ = 931.7 eV. Since the structure of Hg1201 contains only one Cu site, a fluorescence spectrum displays a single maximum (inset to Fig. 2a). The momentum scans were performed by rotating the sample about the axis perpendicular to the scattering plane, and the detector angle was set to $2\theta = 160$deg. The maximum in-plane momentum-transfer component (the projection of the momentum vector onto the CuO$_2$ plane; see Fig. S2) at 931.7 eV was $q = \pm 0.42$ r.l.u.. In the configuration used in the experiment, $K$ was set to zero, and $H$ was coupled to $L$ during the scans. The maximum of the CDW peak was observed at $H = -0.28$ r.l.u., with $L = 1.25$ r.l.u..

RIXS measurements were carried out at the ADRESS beam line [36] of the Swiss Light Source, Paul Scherrer Institut, Villigen PSI, Switzerland. The SAXES spectrometer [37] with fixed scattering angle $2\theta = 130°$ was employed. Similar to the RXD experiment, momentum scans were performed by rotating the sample about the axis perpendicular to the scattering plane. The maximum accessible in-plane component of the momentum transfer was $q = \pm 0.38$ r.l.u.. With $K$ set to 0 the maximum of the CDW peak was observed at $H = -0.28$ r.l.u.. In our scattering geometry these the maximum corresponded to $L = 1.12$ r.l.u.. The energy spectrum of the scattered intensity was analyzed by a grating and recorded with a two-dimensional CCD detector. The energy resolution was 130 meV, as determined from a measurement of the elastic peak of polycrystalline graphite away from the carbon resonance. At each momentum transfer, the energy-loss spectrum was measured five times, each with 10 minute collecting time. The momentum dependence of the quasi-elastic peak (at zero energy loss) was analyzed as a function of momentum transfer. Each energy loss scan was normalized to the total inelastic intensity above the energy-loss of 0.8 eV.

Most of the RIXS measurements were carried out with incident polarization perpendicular to the scattering plane (σ-polarization), and with negative in-plane component of the momentum transfer vector (Fig. S2), as this results in a larger signal. As demonstrated in Ref. 6, the cross section for charge and spin excitations is polarization dependent. For the positive in-plane momenta (Fig. S2), the intensity of spin excitations increases compared to that charge excitations upon switching the polarization from σ to π (polarization vector parallel to the scattering plane). We observed no increase of the intensity as the polarization was changed from σ to π, and therefore can rule out that the observed signal is magnetic.

**Temperature dependence**

The temperature dependence on the CDW peak was studied with the RXD technique. Momentum scans were performed in the temperature range from 4 K to 280 K. The data were collected on increasing the temperature with a constant ramp of 1K/min and every scan corresponded to the temperature change of 1 K. Every four consecutive scans were averaged during the data analysis for further processing. In order to obtain the temperature-dependent contribution to the scattered intensity, the background scattering at 250 K was subtracted from every scan. The background-subtracted data were fit with a Gaussian function. Data collected at a number of temperatures, as well as the temperature dependence of the peak amplitude and width are presented in Fig. S3.

RIXS measurements were performed at 5 temperatures between 10 K and 220 K. The temperature dependences of the CDW peak obtained by the two techniques are in excellent agreement.

**Hard X-ray diffraction measurements**

Hard X-ray diffraction measurements were performed to search for lattice distortions accompanying the CDW on the same sample for which we observed the charge instability with the resonant techniques. The experiment was performed with a six-circle diffractometer at the 6-ID-D beam line of the Advanced Photon Source. The incident photon energy was set to 82.7 keV, energy just below Hg $K$-edge, in order to reduce the fluorescent background. The ratio of the background scattering to the (122) Bragg peak intensity was about $2*10^{-5}$, and the detection limit for incommensurate CDW scattering was ~ $10^{-6}$ of the strong (122) Bragg reflection. No CDW signal was observed. The $1x1x0.5mm^3$ sample was mounted on a cold finger of a standard 'He – closed cycle refrigerator, and the measurement was performed in transmission geometry. An energy dispersive point detector and a two-dimenssional MAR-345 detector were used to collect the diffracted photons. In search of CDW scattering, we explored a wide range of momentum space, using the information gained from the RXD and RIXS experiments. Figure S4a presents a series of theta-2-theta scans along the $H$-direction, with $L$ varying from 5.5 to 9.9 r.l.u. and $K = 0$. The temperature was set to 70 K. Broad thermal diffuse scattering intensity is observed between the shoulders of the Bragg peaks in the neighborhood of integer values of $H$. With the intensity of the (102) reflection 6 orders of magnitude above the background level, there is no indication of the CDW peaks at the $H$ positions (marked by solid vertical lines) where CDW peaks were detected in the resonant scattering experiments. The two-dimensional cuts of the Ewald sphere shown in panels B and C do not reveal any CDW superstructure peaks.

**CDW in YBCO samples from literature**

CDW order was observed in underdoped YBCO samples in the doped charge-carrier concentration range $p$ varying from 0.104 to 0.133. Depending on the hole concentration the samples displayed various types of O-chain order (O-II, O-VII and O-III). A summary of the samples where CDW was detected is presented in Table 1.

| Reference | Sample | O order | $T_C$ (K) | $p$ | $H_{CDW}$ (r.l.u.) | $\xi/a$ |
|---|---|---|---|---|---|---|
| 18 | YBa$_2$Cu$_3$O$_{6.54}$ | II | 58 | 0.104 | 0.32 | 12.4 |
| 6 | YBa$_2$Cu$_3$O$_{6.6}$ | VIII | 61 | 0.11* | 0.31 | 16 |
| 7 | YBa$_2$Cu$_3$O$_{6.67}$ | VIII | 67 | 0.12 | 0.3045 | 20 |
| 18 | YBa$_2$Cu$_3$O$_{6.67}$ | VIII | 67 | 0.123 | 0.305 | 12.3 |
| 6 | YBa$_2$Cu$_3$O$_{6.7}$ | VIII | 69 | 0.13 | 0.304** | 10.6 |
| 17 | YBa$_2$Cu$_3$O$_{6.75}$ | III | 75.2 | 0.133 | 0.3 | 11 |

**Table S1| Summary of CDW in YBCO**

* The hole concentration for this sample has been adopted from the $T_c$ versus $p$ relationship established in Ref. [16]. The hole concentration reported in the original work is $p = 0.12$.

** The CDW modulation wave vector was obtained from a fit of a Gauss function to the published data. The value reported in the original work is $H_{CDW} \sim 0.31$.

**CDW and QO in Hg1201 and YBCO**

To make a connection between the CDW and QO results for Hg1201 and YBCO, the frequency of the oscillations is plotted as a function of the CDW modulation vector $H_{CDW}$ (see Fig. S5 and inset of Fig. 3c). Although the dependence appears to be linear (due to relatively narrow $H_{CDW}$ range) a similarly good fit has been achieved when a quadratic function has been used: the adjusted for the number of predictors $R$-Squared is 0.968 for linear fit and 0.978 for quadratic function of the form $F = a + b/H_{CDW}^2$, where $a = -415(60)$ and $b = 95(6)$. The form of the function connecting $F$ and $H_{CDW}$ has been chosen based on the experimental data in connection with the prediction of the expected $H_{CDW}$ in the limit of vanishing Fermi pocket size ($F = 0$ T). The fit of the quadratic function to the experimental data gives $H_{CDW} = 0.45(2)$ r.l.u. in the limit of vanishing Fermi pocket which should occur for very low hole concentration and result in $H_{CDW}$ spanned between the nodes. This dependence suggests decreasing of the CDW modulation vector with expansion of the FS that gives rise to QO. As frequency is directly proportional to the FS area ($F$ [T]$= \hbar/2\pi e \cdot S$, where; $\hbar$ is Planck's constant, $e$ is elementary charge and $S$ is FS area in Angstroms), the size of the FS expressed as a fraction of BZ ($S \cdot ab/(2\pi)^2$) is a quadratic function of $H_{CDW}$. This is in agreement with scenario in which the electron pockets at nodal points of the BZ give rise to QO with frequency proportional to the area of the electron pocket and with $H_{CDW}$ connecting parts of reconstructed Fermi surface [23]. A simple schematic model is shown in Fig. 3d. An increase of the area of the electron pocket, spanned on the Fermi arc, is related to a decrease of $H_{CDW}$ connecting two parts of FS.

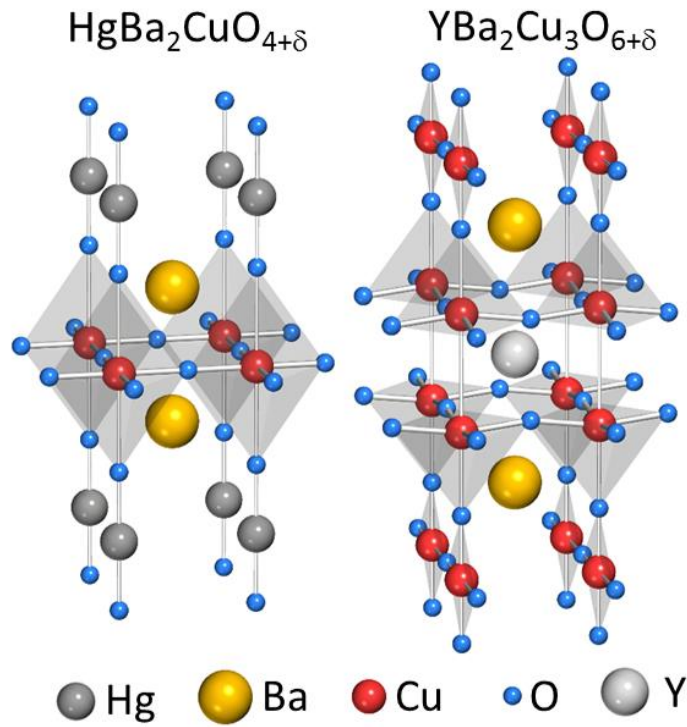

**Figure S1 | Crystal structures of Hg1201 and YBCO.** Hg1201 features tetragonal symmetry (space group P4/mmm) and one $CuO_2$ plane per primitive cell. In the SC doping range, YBCO features lower, orthorhombic symmetry (space group P/mmm) and two $CuO_2$ planes per primitive cell. In both compounds, the charge carrier concentration in the $CuO_2$ planes is controlled by the interstitial O concentration. In YBCO, these O atoms form Cu-O chains along the *b*-axis, with various inter-chain ordering patterns [16]. The interstitial O atoms in Hg1201 reside in the Hg-O planes and do not order in the doping range of interest.

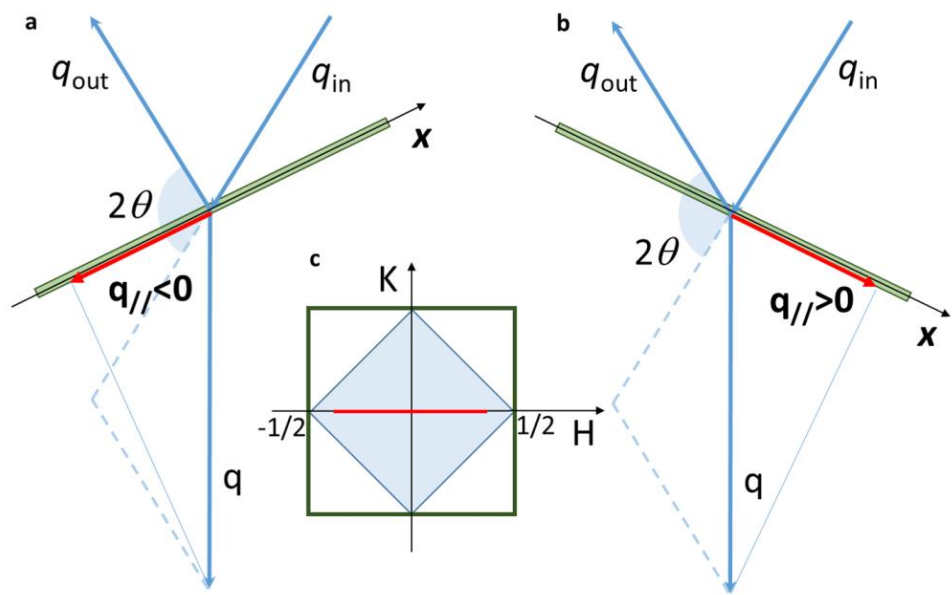

**Figure S2 | Geometry of the resonant X-ray scattering experiments. a**, The red arrow represents an in plane component $q_{//}$ of the momentum transfer vector. $2\theta$ angle was set to 160 deg in RXD and 130 deg in RIXS experiment. Here $q_{//}$ is negative. **b** As in a but for $q_{//}>0$ configuration. **c** Structural (green square) and magnetic (blue square) Brillouin zone in H-K plane. The red bar represents the accessible $q_{//}$ range: ±0.42 r.l.u. in RXD and ±0.38 in RIXS experiment.

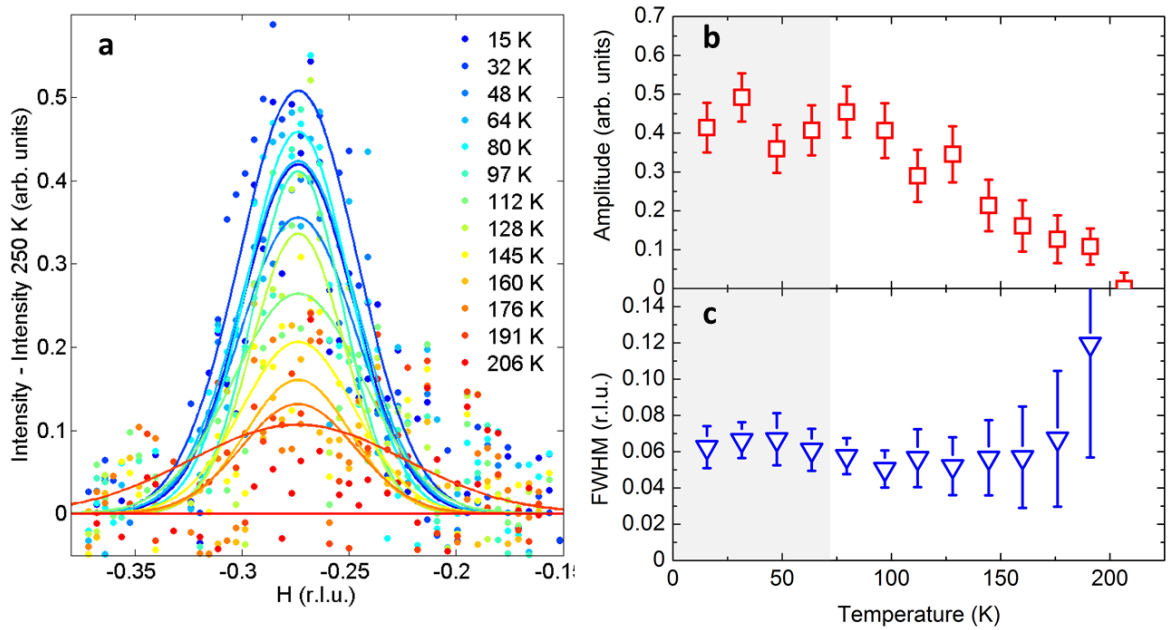

**Figure S3 | Temperature dependence of the CDW peak observed by RXD. a**, Momentum scans at several temperatures, with the spectrum collected at 250 K subtracted. The data were fit with a Gaussian function with a common $H_{CDW}$ position. **b**, Temperature dependence of the CDW peak amplitude determined from the fits presented in panel (a). **c**, FWHM of the Gaussian fit to the CDW peak plotted as a function of temperature. The correlation length, defined from the FWHM of the Gaussian as $\xi/a = (1/2\pi/\text{FWHM})$, is expressed in the lattice units and presented in Fig. 1e.

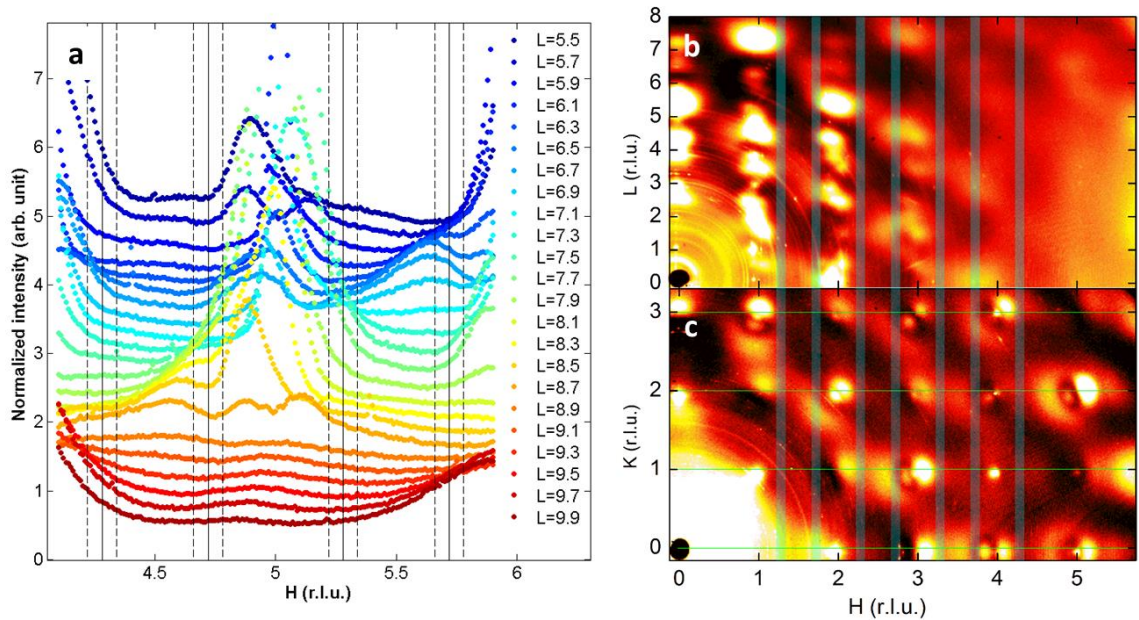

**Figure S4 | Diffuse hard-X-ray scattering for Hg1201 sample with $T_c$ = 72 K, collected at 70 K. a**, $H$-scans performed with a point detector at many $L$ values, with $K = 0$. Solid vertical lines indicate the momentum position of the CDW peak as observed with resonant X-ray scattering techniques, and the dashed lines mark the FWHM. The main contribution to the observed intensity is thermal diffuse scattering. **b**, Two-dimensional cuts of the Ewald sphere in the $H$-$L$ plane. **c**, Two-dimensional cuts of the Ewald sphere in the $H$-$K$ plane. The position of the shaded stripes in b and c corresponds to the expected $H_{CDW}$ value. The width of the stripes corresponds to the FWHM expected from RXS measurements. The diffraction patterns have been recorded with two-dimensional MAR345 detector.

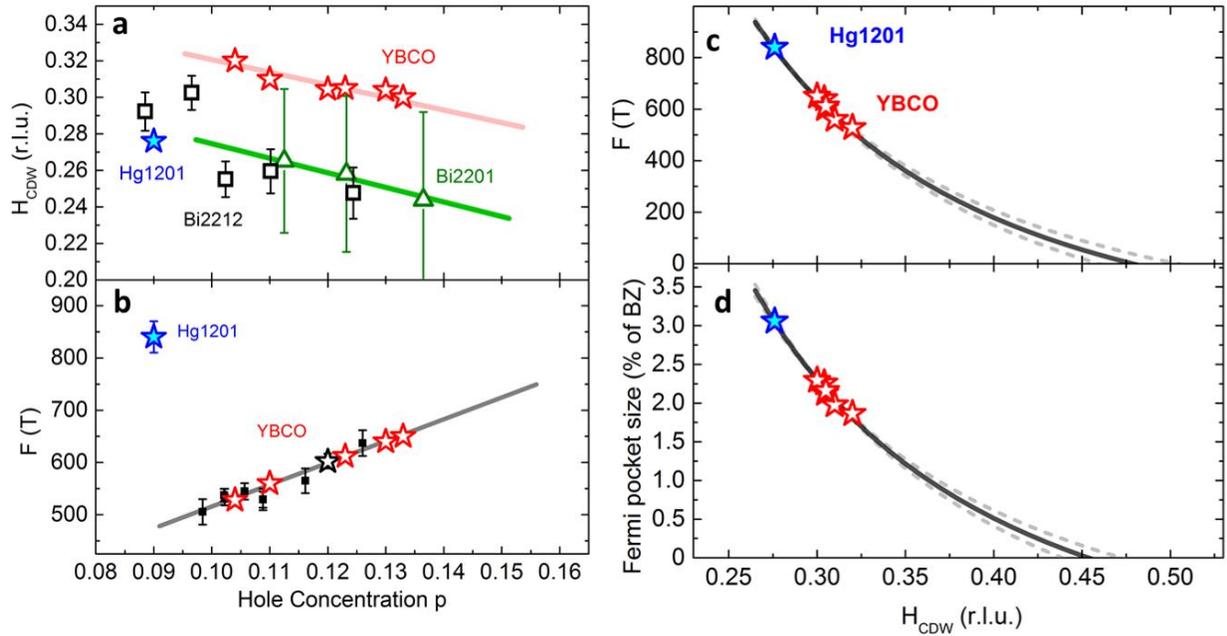

**Figure S5 | CDW wave vector and quantum oscillation period. a**, $H_{CDW}$ for Hg1201 (this work), YBCO [4,24], Bi2201 [27] and Bi2212 [30] as a function of hole concentration. This dependence is linear for a given compound. **b**, Frequency of Shubnikov-de-Haas oscillations (QO) revealed by high magnetic fields at low temperatures (black squares). The grey solid line is a linear fit to the experimental data. The black star indicates a doping level for which both CDW and QO have been measured in YBCO. Red stars represent the QO frequencies estimated form the linear fit for the doping levels at which only CDW order (but not QO) was studied. **c**, QO frequency as a function of the $H_{CDW}$ for Hg1201 (blue star) and YBCO (red stars). The solid line represents a quadratic fit to the experimental data and the dashed lines show the fit error range. **d**, Fermi pocket size, expressed as a fraction of the first Brillouin zone (BZ) versus $H_{CDW}$ for Hg1201 (blue) and YBCO (red), with doping as intrinsic parameter. The solid line represents a quadratic fit, which extrapolates to $H_{CDW} = 0.45(2)$ r.l.u. in the limit of vanishing pocket size. The dashed lines show the fit error range.